\documentclass[journal]{IEEEtran}
\PassOptionsToPackage{bookmarks=false}{hyperref}
\usepackage{xurl}

\usepackage{etex}
\usepackage[normalem]{ulem}

\usepackage{graphicx,amsmath,amssymb,color,url,algorithmic,float,multirow,comment}

\usepackage{placeins}
\usepackage[ruled]{algorithm2e}
\usepackage{arydshln}
\usepackage{xspace}
\usepackage{chngpage} 
\usepackage[table]{xcolor}
\usepackage{epstopdf}
\usepackage{gensymb}
\usepackage{threeparttable}
\usepackage{booktabs, dcolumn}
\usepackage{caption}
\usepackage{subcaption}
\usepackage{xcolor}
\usepackage[colorlinks]{hyperref}
\usepackage{filecontents}

\usepackage{cite}
\usepackage[numbers,sort&compress]{natbib}
\usepackage{fancyhdr}

\begin{document}
\bstctlcite{IEEEexample:BSTcontrol}

\title{When Everyday Devices Become Weapons: A Closer Look at the Pager and Walkie-talkie Attacks}

\author{\IEEEauthorblockN{Pantha Protim Sarker, Upoma Das, Nitin Varshney, Shang Shi, Akshay Kulkarni, Farimah Farahmandi, and Mark Tehranipoor}

\IEEEauthorblockA{\textit{Department of Electrical and Computer Engineering, University of Florida, Gainesville, FL, USA}\\
\{psarker, upoma.das, nitinvarshney1, shang.shi, kulkarniakshay\}@ufl.edu\\
\{farimah, tehranipoor\}@ece.ufl.edu
}}

%\markboth{Future Hardware Security Research Series}{}

\maketitle
\begin{abstract} 
Battery-powered technologies like pagers and walkie-talkies have long been integral to civilian and military operations. However, the potential for such everyday devices to be weaponized has largely been underestimated in the realm of cybersecurity. In September 2024, Lebanon experienced a series of unprecedented, coordinated explosions triggered through compromised pagers and walkie-talkies, creating a new category of attack in the domain of cyber-physical warfare. This attack not only disrupted critical communication networks but also resulted in injuries, loss of life, and exposed significant national security vulnerabilities, prompting governments and organizations worldwide to reevaluate their cybersecurity frameworks. This article provides an in-depth investigation into the infamous Pager and Walkie-Talkie attacks, analyzing both technical and non-technical dimensions. Furthermore, the study extends its scope to explore vulnerabilities in other battery-powered infrastructures, such as battery management systems, highlighting their potential exploitation. Existing prevention and detection techniques are reviewed, with an emphasis on their limitations and the challenges they face in addressing emerging threats. Finally, the article discusses emerging methodologies, particularly focusing on the role of physical inspection, as a critical component of future security measures. This research aims to provide actionable insights to bolster the resilience of cyber-physical systems in an increasingly interconnected world.
\end{abstract}


\section{Introduction}\label{sec:intro}

\subsection{Hardware Attacks: Risks and Realities}
In recent years, hardware-based attacks have become a prominent concern in cybersecurity research, especially as cyber-physical systems increasingly merge with digital and connected infrastructure. %Hardware attacks are distinctive in that they bypass many software-level defenses by targeting the physical properties, architecture, or embedded functionalities of a device, leveraging these attributes to create vectors for unauthorized access, data exfiltration, or even sabotage. 
Hardware attacks stand out because they bypass software-level defenses by exploiting a device's physical properties, architecture, or embedded functions, enabling unauthorized access, data theft, or sabotage. As societies integrate more critical systems into the Internet of Things (IoT) and rely on embedded systems, hardware security vulnerabilities have gained serious attention~\cite{hu2020overview, rahman2018physical}. The risk inherent in global supply chains and the need for advanced security measures to prevent unauthorized tampering have been highlighted by high-profile incidents involving modified hardware components~\cite{mehta2020big,guin2014counterfeit}.

Hardware attacks often exploit the fact that devices like processors, memory, and communications chips can be covertly altered to introduce backdoors, unauthorized data collection capabilities, or even destructive functionalities. This approach contrasts with traditional cyberattacks that typically require access to network or software interfaces; In hardware attacks, physical manipulation, electromagnetic (EM) signals, and supply chain infiltration are key factors. An example of this is the use of malicious modifications (hardware trojans) in devices at the manufacturing stage, which allows adversaries to plant vulnerabilities before devices even reach their intended users~\cite{xiao2016hardware,bhunia2018hardware}. These modifications can create backdoors, leak data, or alter functionality to degrade performance or enable sabotage. Once in use, these tampered components can be controlled remotely or programmed to act maliciously under specific conditions. Unlike software vulnerabilities, hardware Trojans are exceptionally difficult to detect once embedded, often requiring advanced techniques such as side-channel analysis or physical inspection.

Globalization has played a significant role in increasing hardware vulnerabilities~\cite{tehranipoor2011introduction,tehranipoor2016survey,guin2014counterfeit}. As companies outsource manufacturing to various regions with differing levels of oversight and security practices, there is an increased risk of unauthorized alterations during the manufacturing and assembly phases. In particular, adversaries may target foundries and component suppliers to introduce vulnerabilities into a product before it reaches the market. The complexity and number of steps in the supply chain make comprehensive verification challenging, which leads to vulnerabilities at multiple stages. For instance, if a processor is produced at one facility, packaged at another, and finally assembled into a printed circuit board (PCB) in a third country, each step could provide an opening for unauthorized tampering or manipulation.

\subsection{The Pager and Walkie-Talkie Attacks}

\begin{figure}[htbp]
    \centering
    \includegraphics[width=\columnwidth]{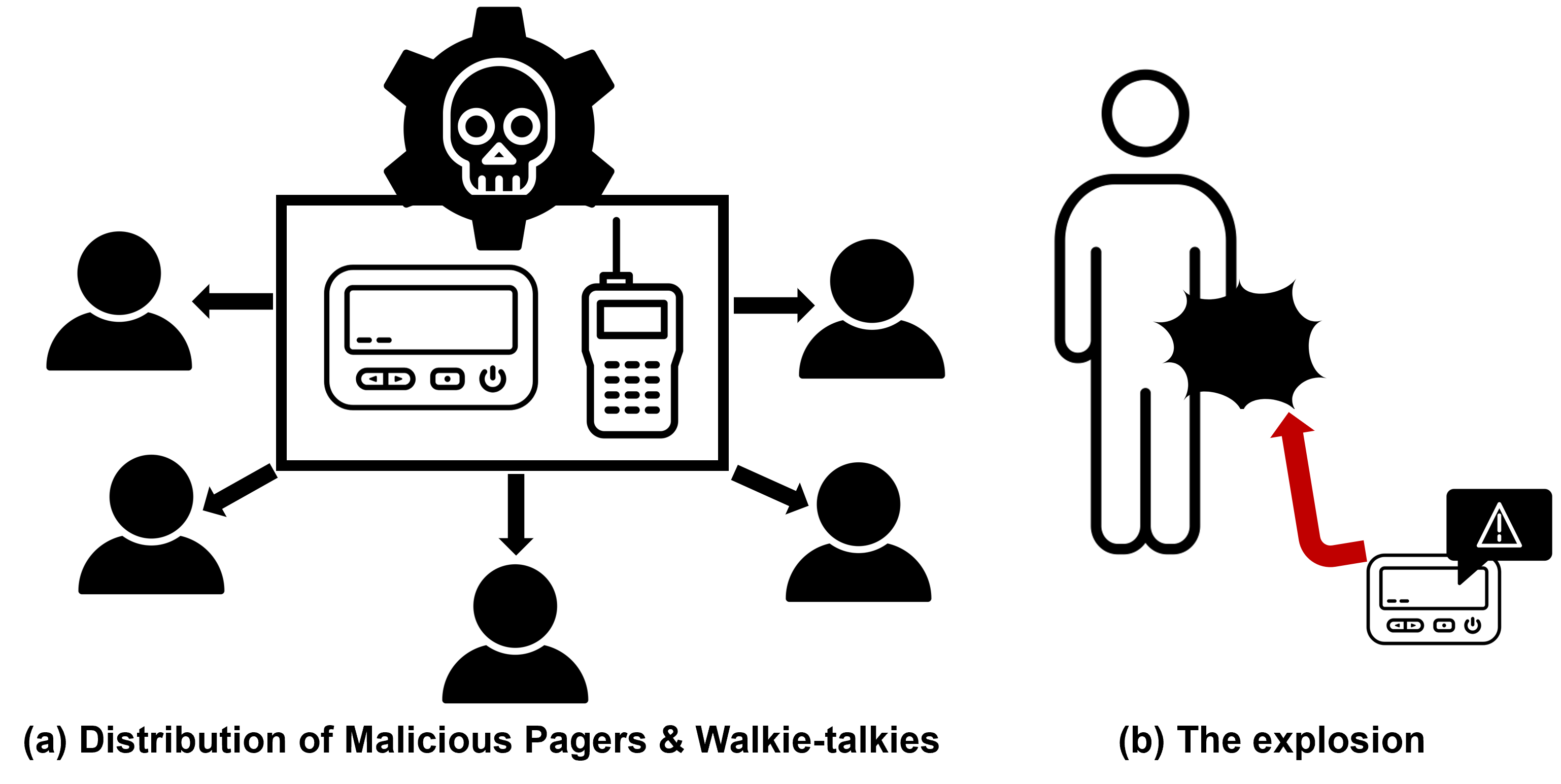}
    \caption{A brief overview of the pager and walkie-talkie attacks}
    \label{fig_overview}
\end{figure}

\subsubsection{Attack Background}
A recent and widely known example of a hardware-based exploit is the pager and walkie-talkie attack (see Fig.~\ref{fig_overview}), which occurred in Lebanon in September 2024~\cite{wiki_1}. On September 17, thousands of pagers exploded simultaneously in Hezbollah strongholds, followed by a second attack the next day involving weaponized walkie-talkies. The coordinated explosions killed 42 people and injured over 3,500~\cite{Reuters_1}. This attack targeted high-profile Hezbollah operatives using malicious communication devices that were suspected to be rigged with hidden explosives. The tampered devices not only caused physical damage and fatalities but also disrupted secure communication channels, showcasing the devastating potential of hardware-based attacks. %It is likely that pagers and walkie-talkies, both of which use standardized components, were earmarked because of their widespread adoption and predictable supply chains. This incident highlights the growing sophistication of hardware threats and their ability to exploit weaknesses within the global supply chain.
Pagers and walkie-talkies, with their standardized components and predictable supply chains, were likely targeted for their widespread use. This incident underscores the increasing sophistication of hardware threats and their exploitation of global supply chain vulnerabilities.

Hezbollah reportedly sourced the pagers directly from a Taiwanese company named Gold Apollo~\cite{wiki_1}, making this company a critical link in the supply chain exploited during the attack. Gold Apollo, founded in 1995, specializes in pagers and associated communication technology~\cite{wiki_2}. Despite being a relatively small company with approximately 40 employees, Gold Apollo holds a significant position in the pager market, particularly in North America and Europe. Its products are also widely used in other sectors, including critical ones like security and emergency services~\cite{wiki_2}. This widespread adoption makes their devices attractive targets for adversaries seeking to exploit trust in well-established brands. In addition, Gold Apollo had a licensing agreement with BAC Consulting, a Hungarian company, that provided a cover for the introduction of tampered devices~\cite{wiki_1}. This arrangement allowed the attackers to mask their involvement and facilitated the manufacturing and/or distribution of the malicious pagers. By implanting explosive components or triggering mechanisms into devices, the attackers weaponized everyday communication tools, converting them into covert assassination devices.

Hardware attacks, such as those executed through compromised pagers and walkie-talkies, have far-reaching implications. Unlike software-based threats, which can often be mitigated through updates and patches, hardware compromises are much harder to detect and rectify, especially after deployment. These attacks not only pose a direct physical threat but also erode confidence in the reliability of communication infrastructure. As hardware tampering becomes a more viable avenue for adversaries, it underscores the need for improved security practices, including enhanced traceability, rigorous supply chain audits, and the adoption of tamper-resistant designs.

\subsubsection{Attack Timeline} %\textcolor{red}{(Pantha)}} % alternate title
\begin{figure}[http]
    \centering
    \includegraphics[width=\columnwidth]{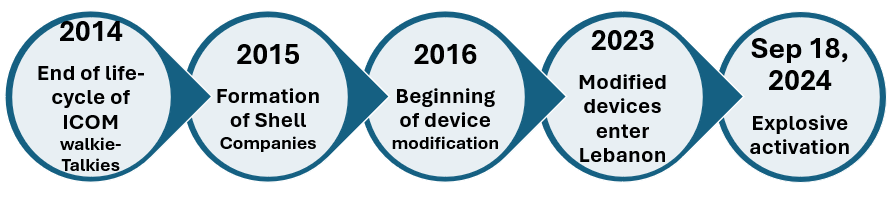}
    \caption{Timeline of planning and execution of the walkie-talkie attacks in Lebanon}
    \label{fig_timeline}
\end{figure}

Investigations revealed long complex planning to deliver modified pagers and walkie-talkies into Hezbollah-controlled regions~\cite{bbc_1}. %The operation is suspected to have been orchestrated by Israeli intelligence agencies, including Mossad and Shin Bet, with the support of the Israeli Defense Forces (IDF)~\cite{CNN_1}.%
The attacks are part of a covert operation involving multiple entities across different countries~\cite{CNN_1}. Multiple shell companies (i.e., Norta Global, BAK holding etc.) played a central role in distributing these devices under the guise of legitimate brands like Gold Apollo~\cite{Reuters_1}. Financial transactions between these entities obscured the origin of the tampered devices~\cite{cradle}. Reports suggest that these devices were inserted into Lebanon starting in 2015, approximately a year after their original production had been discontinued. The timeline for walkie-talkie attack is summarized in Fig.~\ref{fig_timeline}. The timeline indicates a long-term surveillance and sabotage strategy, starting with the acquisition of discontinued ICOM walkie-talkies in 2014 and the establishment of shell companies in 2015. Device modifications started in 2016, with the tampered devices entering Lebanon in 2023 and ultimately being activated during the September 18, 2024 explosions. The operation demonstrates detailed planning regarding supply chain infiltration and the technical modifications made to the devices.

\subsection{Techinical Overview of Pagers and Walkie-Talkies}% (\textcolor{red}{Pantha})}
To better understand the attack execution, it is essential to examine the technical details of pagers and walkie-talkies first. Pagers and walkie-talkies are considered low-tech devices, making them less prone to digital tracking and data interception~\cite{enck2014taintdroid, louk2014analysis}. %In contrast, modern smartphones offer extensive communication capabilities but also generate a wealth of trackable information, such as location data, communication logs, and usage patterns, which can be exploited for surveillance~\cite{enck2014taintdroid, louk2014analysis}.

\subsubsection{Pager Architecture}
A pager, also known as a beeper, is a wireless communication device that receives and displays messages. It is primarily a one-way communication device. Pagers are widely used in scenarios where instant and reliable communication is critical, such as in healthcare, emergency services, and industrial environments~\cite{quan2013s}. Pagers function by capturing radio signals transmitted over dedicated frequencies or networks, which are typically assigned by regulatory bodies to minimize interference~\cite{patel2010resident}. A microcontroller within the pager decodes these incoming signals and verifies if they match the device’s unique address or identification code. Once matched, the messages are stored in memory and can be displayed or retrieved later. Paging communication relies on established protocols such as POCSAG, a widely-used standard that operates on low-frequency bands to enable long-range communication~\cite{jacques2006evaluation}. FLEX~\cite{xiaojian2000novel}, an advanced protocol, enhances this capability by offering higher data rates and improved message reliability. Pagers generally operate within UHF or VHF bands, providing wide coverage and reliable signal penetration in buildings~\cite{clancy1998designing}.

%\subsection{Technical Details of Walkie-Talkie (\textcolor{red}{Pantha})}

\subsubsection{Walkie-Talkie Architecture}
A walkie-talkie, also known as a two-way radio, is a handheld communication device that supports bidirectional voice communication. Unlike pagers, walkie-talkies allow real-time conversations, making them ideal for team coordination in professional, recreational, or emergency contexts~\cite{bandyopadhyay2010wireless}. Walkie-talkies integrate both a transmitter and receiver into a single unit. The device’s key feature, the Push-to-Talk (PTT) button, switches between transmitting and receiving modes. An external or internal antenna facilitates the transmission and reception of radio waves. Powered by rechargeable lithium-ion batteries for extended use, modern walkie-talkies incorporate processors that handle signal modulation, demodulation, and encryption when supported. These devices operate on specific frequency bands, such as Very High Frequency (VHF), which offers long-range communication in open areas, and Ultra-High Frequency (UHF), which provides better penetration in urban or indoor environments~\cite{rappaport2024wireless}. Communication is typically based on FM modulation for stable voice transmission or digital modulation for enhanced audio quality and encryption. Advanced features include selective calling systems like Continuous Tone-Coded Squelch System (CTCSS), which filters radio traffic to allow only designated signals, and Dual Tone Multi-Frequency (DTMF), enabling paging and individual call addressing~\cite{rappaport2024wireless}.

%Use of Pagers
\subsubsection{Hezbollah's Shift to Pager and Walkie-Talkies}
The adoption of pagers by Hezbollah represents an attempt to avoid the vulnerabilities associated with smartphone-based digital surveillance~\cite{enck2014taintdroid, louk2014analysis}. Unlike smartphones, pagers operate on simpler technology with limited features, making them harder for adversaries to infiltrate or tamper with remotely. Hezbollah members have used pagers as backup communication devices for years; however, %Hassan Nasrallah,%
Hezbollah's leadership, ordered operatives to replace smartphones with pagers in early 2024, which resulted in a surge in pager usage. Unfortunately, Hezbollah was unaware that in many of these new devices, the batteries had been rigged to malfunction when activated by a specific trigger.

The specific pager models used by Hezbollah were reportedly sourced from Taiwan’s Gold Apollo Technology, a manufacturer known for producing high-end pagers used in specialized fields like healthcare. The devices met Hezbollah’s security criteria by offering long battery life and the ability to recharge, key features for reliable covert communication. Nonetheless, it is believed that these devices were intercepted by Israeli operatives during distribution and modified to carry small amounts of high-powered explosives. The modifications were allegedly facilitated by BAC Consulting Kft, a Hungarian intermediary working under a license agreement with Gold Apollo, and possibly connected to shell companies used by Israeli intelligence agencies.

The use of walkie-talkies, likewise, followed a similar pattern. Previously, Israeli agencies planted walkie-talkies in Lebanon for surveillance purposes, but the recent attacks signal a shift toward offensive applications, where the device serves as both a communication tool and a potential weapon. This strategic dual-use of hardware devices, capable of functioning as both a communication medium and a targeted explosive, underscores the pressing need for secure hardware designs and trusted sourcing in critical operational environments.

The pager and walkie-talkie incidents underscore the growing necessity of secure hardware in modern warfare and intelligence, where modified commercial electronics can covertly become weapons. As supply chains globalize, the potential for hidden modifications, especially in regions with mixed trusted and untrusted suppliers, introduces significant risks. This incident reflects a broader trend toward hardware-based vulnerabilities, emphasizing the need for enhanced inspection techniques, verification protocols, and security assurances across all components in communication systems.

\section{The Impact } \label{sec:impact}
%What was the impact of the attack

%had profound consequences for both the organization and the people in Lebanon.
The pager and walkie-talkie attacks had profound consequences for the organization, the people in Lebanon, and the broader global community. On September 17th, the detonation of pagers resulted in at least 12 fatalities, including two young children, and injuring over 2,750 people~\cite{wiki_1}. The following day, a second wave of explosions ripped through walkie-talkies, claiming the lives of at least 30 individuals and leaving more than 750 wounded. The attacks, which spanned across Lebanon and even reached into Syria, overwhelmed medical facilities and caused chaos throughout the region. The total death toll is estimated to be at least 42, with the vast majority of casualties occurring within Lebanon~\cite{wiki_1}. These explosions also disrupted critical communication networks, resulting in operational setbacks and heightened fears about the security of widely used communication devices, affecting both civilians and local organizations. This attack also has a significant global impact. It underscores the significant and far-reaching consequences of hardware vulnerabilities. It demonstrates the use of embedded systems as weapons to target organizations and industries globally, emphasizing the urgent need for robust hardware security measures. This section explores the long-term effects of such attacks on private companies, governments, and the global tech ecosystem, demonstrating how hardware compromises can ripple across industries and borders.

\subsection{Impacted Companies}
Gold Apollo, the Taiwanese company implicated as the source of the compromised devices, faced significant reputational damage. Although no direct evidence linked the company to intentional misconduct, its role in the global supply chain made it a focal point of scrutiny. Following the incident, Gold Apollo's stock value reportedly dropped by 15\% within weeks, as clients reevaluated their reliance on its products~\cite{bloomberg_1}. Taiwanese authorities have searched multiple locations and questioned individuals associated with Gold Apollo. The Hungarian company 'BAC Consulting', which was licensed to use Gold Apollo's brand, also faced scrutiny and investigation~\cite{wiki_1,nbc_1}. Additionally, the incident raised concerns about the involvement of other intermediaries and shell companies, complicating the situation further~\cite{nbc_1,cnn_2}.

The attack also had broader implications for the electronics and telecommunications industries, as it highlighted vulnerabilities in the supply chain and the potential for misuse of technology. Telecommunications and electronics companies worldwide faced a surge in demand for supply chain audits and secure sourcing. Taiwan's tech industry, which plays a crucial role in global electronics, has faced increased scrutiny and pressure to ensure the integrity of its products~\cite{aljazeera_1}. Manufacturers of similar devices—pagers, walkie-talkies, and other embedded communication systems—reported losses as industries hesitated to procure devices from suppliers with insufficient traceability. These events served as a wake-up call, highlighting the vulnerabilities inherent in globalized production networks and prompting many firms to reexamine their procurement practices and quality assurance protocols.

\subsection{Impact on Governments and National Security}

Governments worldwide, including the United States, closely monitored the attack and its aftermath. U.S. intelligence agencies expressed concerns about potential risks to their own supply chains, particularly in sectors involving critical communication systems~\cite{aljazeera_1,npr_1}. The incident raised alarms over imported technology used in sensitive operations, prompting calls for increased investment in domestic manufacturing and enhanced security measures for hardware imports.

Lebanon’s government faced additional strain as it dealt with the dual challenges of humanitarian aid for the wounded and securing its own communication infrastructure~\cite{wiki_1}. Regional governments in the Middle East also began reassessing the security of their equipment, leading to policy shifts and an uptick in regulatory scrutiny of imported communication devices.

\subsection{Broader Global Implications}
The ripple effects extended beyond the Middle East, affecting industries and nations globally. In the U.S., for instance, the Department of Defense (DoD) and private-sector contractors supplying communication systems initiated emergency reviews of their supply chains~\cite{wiki_1}. Similar responses occurred in Europe, where industries dependent on embedded communication systems—such as telecommunications, automotive, and aerospace—began implementing stricter procurement standards.

The attack highlighted vulnerabilities in the horizontal business models underpinning globalized production. With components sourced from multiple regions and assembled by different contractors, the complexity of the supply chain provided adversaries numerous points of entry~\cite{wiki_1}. The incident also underscored the need for robust traceability, particularly for widely adopted devices like pagers and walkie-talkies, which are often viewed as mundane yet indispensable tools across industries.
\section{Different perspective on the attack} %(\textcolor{red}{Pantha})} 
\label{sec:theories}
After the pager and walkie-talkie attack, several theories emerged from experts and sources on the attack and its execution. These theories range from technical hypotheses about how the devices were modified, how the explosion was carried out, and what were the motives behind the attacks. Here are some prominent theories proposed in the aftermath.

\subsection{Explosive Concealed in the Battery}
One widely discussed theory suggests that these devices were intentionally equipped with explosive material. In this case, The explosives were concealed within the battery packs of the devices. The primary explosive, Pentaerythritol Tetranitrate (PETN), popular as RDX, was embedded inside the Li-ion battery casing. PETN, being highly powerful yet chemically stable, made the battery an ideal location for concealment while maintaining the device’s functional appearance while embedding a lethal charge~\cite{Reuters_1}. This approach effectively combined concealment, reliability, and the ability to evade conventional security checks. However, the reduced battery capacity could have raised suspicion during detailed inspections.

\subsection{Explosives Concealed Elsewhere in the Device}
An alternate theory posits that the explosive material could have been concealed in compartments other than the battery, such as the casing or unused internal spaces. This would have allowed the battery to retain its full functional capacity and thus would have reduced suspicion. In larger devices like walkie-talkies, the casing offers more space to accommodate a larger explosive payload, which could result in a more devastating impact~\cite{crypto_m_1}. However, this approach presents greater risks of discovery during assembly or operational use and is less plausible for compact devices like pagers, which lack substantial internal compartments.

\subsection{No Explosive, Only Thermal Runaway}
Another perspective suggests that the devices contained no explosives but instead relied on inducing a thermal runaway in the battery cells~\cite{yin2023review}. This phenomenon, caused by overheating or internal short-circuiting, could lead to an explosion and fire~\cite{tran2022review}. The triggering mechanism would involve remotely overloading the battery or introducing an internal spark. While this theory offers a simpler design and avoids the use of explosive materials, it lacks the precision and control required for such a targeted attack. Moreover, the destructive potential of thermal runaway is significantly lower compared to an explosive like PETN, making this theory less likely.

\subsection{Remote Triggering Mechanism}
Experts have also theorized that the explosions were initiated through remote triggering mechanisms, leveraging the devices' existing communication features. Technologies like Continuous Tone-Coded Squelch System (CTCSS) or Dual-Tone Multi-Frequency (DTMF) coding could have been used to send specific signals to the devices, activating the detonators. The detailed mechanism will be discussed in subsection~\ref{subsec_trigger}. These technologies could have been used to send specific signals to the devices, activating the detonators upon receipt of a unique code~\cite{vao}. This approach offers precise timing and coordination for controlled activation of the devices while maintaining their functionality as communication tools. However, this method requires sophisticated programming and access to secure communication networks, making it a highly technical operation that only a well-resourced entity could execute~\cite{crypto_m_1}.

\subsection{Proximity-Based or Environmental Triggers}
Another theory involves proximity-based or environmental triggers. It suggests that the devices were equipped with sensors or mechanisms that activated the explosives when specific conditions were met~\cite{purcaru2017study}. For example, a vibration, a change in temperature, or proximity to a particular electromagnetic field could have served as the triggering factor. %This method would bypass the need for remote signaling, making the devices independent of external communication networks and reducing the chances of detection through network monitoring. However, this theory is less widely accepted because environmental triggers are inherently unpredictable and may not guarantee the precise activation required for such a carefully orchestrated operation.
This method avoids remote signaling and does not rely on communication networks, reducing the risk of detection. However, it is less accepted because environmental triggers are unpredictable and may not ensure precise activation to start a well-coordinated attack.

\subsection{Explosives as a Primary Function}
According to the experts, there are varying theories about the intended purpose of the devices. The prominent theory suggests that the primary intent of the devices was to act as covert improvised explosive devices (IEDs). The %dual-use functionality 
architecture and communication protocol of pagers and walkie-talkies provided a perfect disguise for integrating explosives. This theory states that the explosions were planned well in advance, with the attackers relying on the devices’ appearance to infiltrate and embed them deep within Hezbollah’s operations~\cite{bbc_2}. By leveraging advanced triggering mechanisms, the perpetrators ensured precise timing for the explosions.

\subsection{Hybrid Purpose: Espionage Followed by Targeted Sabotage}
Some experts are also suggesting they were designed for more than just explosions. One idea is that the devices were primarily intended for surveillance and communication monitoring~\cite{bbc_1}. Modified to blend seamlessly into Hezbollah’s infrastructure, these pagers and walkie-talkies may have allowed attackers to intercept messages, track movements, and gather valuable intelligence over an extended period. In this scenario, the explosive modifications could have been a secondary feature, perhaps intended as a backup plan to neutralize the devices or target specific individuals if needed. While this theory is highly plausible, no concrete evidence has yet emerged from the analysis of the exploded PCBs of the devices.

Each theory provides valuable insights into the motivations and methods behind the attacks. They range from intelligence gathering to supply chain infiltration. A summary of the strengths and weaknesses of these theories is presented in Table~\ref{tab_summary_theories}.

\begin{table*}[ht]
\centering
\caption{Summary Table of Theories on the Attacks}
\label{tab_summary_theories}
\begin{tabular}{|l|l|l|p{3.6cm}|p{3.6cm}|}
\hline
\textbf{Category}                 & \textbf{Theory}                       & \textbf{Plausibility}      & \textbf{Strengths}                                             & \textbf{Weaknesses}                                      \\ \hline
\multirow{3}{*}{\textbf{Explosive Integration}} & Explosive Concealed in Battery        & High                  & Concealment, reliability, evades detection               & Reduced battery capacity could hint at tampering        \\ \cline{2-5} 
                                   & Explosives Concealed Elsewhere        & Moderate              & Maintains full battery functionality, more space in larger devices & Higher risk of discovery, less plausible for pagers    \\ \cline{2-5} 
                                   & No Explosive, Only Thermal Runaway    & Low                   & Simpler design, avoids explosive detection               & Uncontrolled outcome, limited destructive power         \\ \hline
\multirow{2}{*}{\textbf{Triggering Mechanism}}  & Remote Triggering Mechanism           & High                  & Precision, control, aligns with communication capabilities & Requires advanced technical expertise                   \\ \cline{2-5} 
                                   & Proximity/Environmental Triggers      & Low                   & Independent of communication networks, avoids detection  & Unpredictable and lacks precision                       \\ \hline
\multirow{2}{*}{\textbf{Intended Use}}              & Explosives as Primary Function        & Moderate              & Straightforward IED design, premeditated attack          & Overlooks potential dual-use functionality              \\ \cline{2-5} 
                                   & Hybrid Purpose: Espionage/Sabotage    & High                  & Explains long-term use, dual-purpose functionality       & No concrete evidence from PCBs                          \\ \hline
\end{tabular}
\end{table*}

\section{Execution of the Attack}
\label{execution}
After evaluating the expert theories summarized in Table~\ref{tab_summary_theories}, we selected the most plausible explanation based on the available evidence and logical reasoning. Using the chosen theory as a foundation, we will now describe in detail how the attacks were likely carried out.

\subsection{Technical Modifications of the Devices and Systems}
The devices used in the attacks were heavily modified to function as both communication tools and explosive mechanisms. %According to the expert, 
In this most plausible explanation, the pagers and walkie-talkies featured concealed explosive components within the battery ~\cite{Reuters_1}.

\subsubsection{Embedded Explosives within Battery Packs}
In this scenario, the explosive devices adhered to the basic principles of an improvised explosive devices (IED), which typically consists of five key components: a switch, power source, detonator, explosive charge, and container~\cite{dhs_gov}. When using pagers like the Gold Apollo AR-924 or handheld radios like the ICOM IC-V82, three of these components—switch, power source, and container—were inherently present, significantly simplifying the modification process. The modifications introduced the remaining elements: the PETN as the explosive charge and a spark-based mechanism as the detonator.

The battery was intelligently modified to integrate both explosives and a detonator. The primary explosive used, Pentaerythritol Tetranitrate (PETN), is highly flammable yet chemically stable, requiring a slightly unstable detonator that could be triggered by a small spark, making battery a very smart choice for these attacks. Each component was carefully arranged to conceal explosive materials while maintaining some of the battery’s functional characteristics. The battery pack consisted of three components:

\begin{figure}[htbp]
\centerline{\includegraphics[width=\columnwidth]{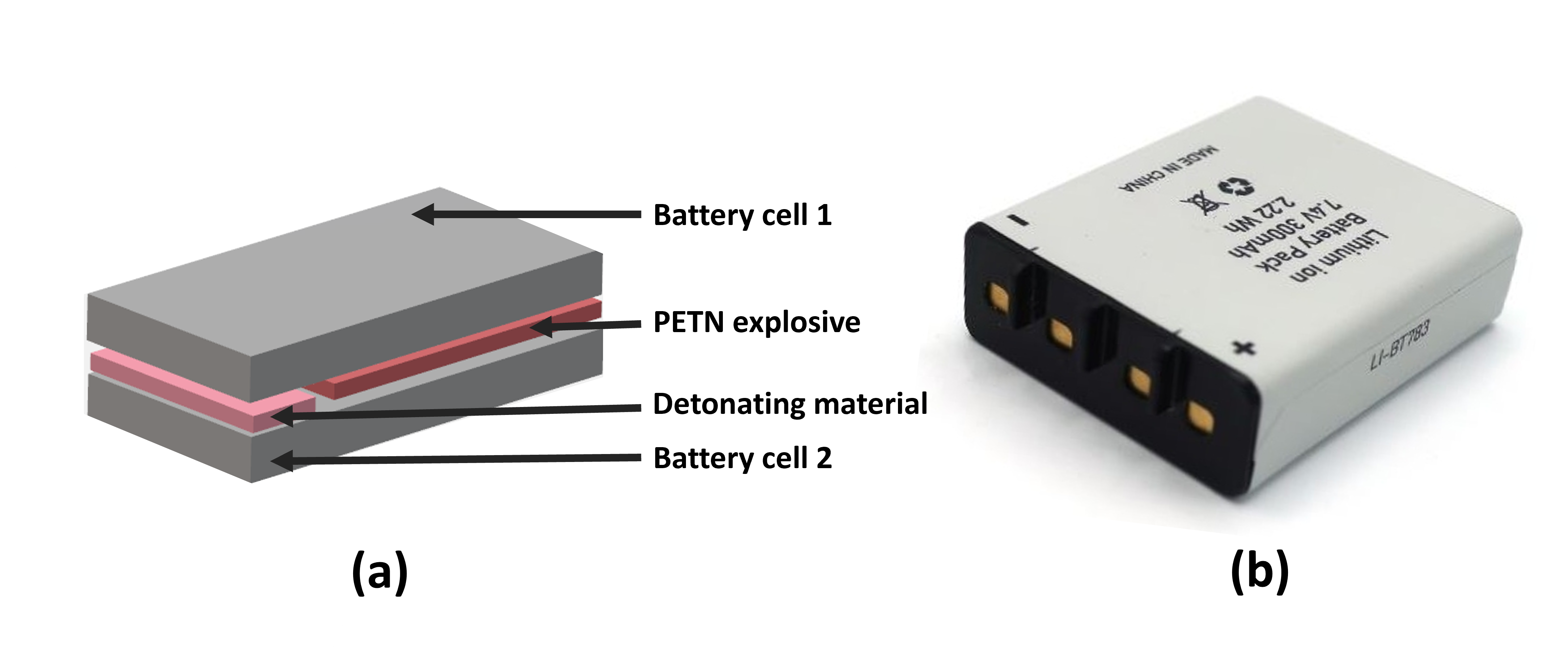}}
\caption{Battery construction with PETN explosives. (a) Two Lithium-ion cells sandwich a sheet of plastic explosive and detonator, (b) Li-BT783 battery pack~\cite{Reuters_1}}
\label{fig:battery}
\end{figure}

\begin{itemize}
    \item \textit{PETN Explosive Sheet:} The central element of the concealed explosive mechanism was a thin, square sheet of PETN, weighing approximately six grams~\cite{conversation}. As shown in Fig.~\ref{fig:battery}, this plastic explosive was sandwiched between two rectangular battery cells, maintaining the appearance of a standard battery pack while embedding a highly destructive charge. Later, battery experts noted that the modified battery packs had significantly reduced capacity due to the space occupied by the explosive material and wrapping. For example, a standard 35-gram battery of this size would be expected to provide approximately 8.75 watt-hours (Wh) of energy, but the modified battery delivered only 2.22 Wh~\cite{Reuters_1}. This discrepancy, along with the unaccounted mass, was consistent with the integration of PETN and other materials.
    \item \textit{Detonator Material:} Instead of using a conventional metallic detonator,  which would have been easily detectable by X-ray scanners, the design incorporated a non-metallic strip of highly flammable material placed between the battery cells. This strip functioned as a spark-sensitive detonator, specifically selected for its low detection profile and its capability to reliably ignite the PETN under controlled conditions. When the devices were received in February 2024, Hezbollah subjected them to airport security scanner tests for explosives~\cite{DW}. However, the modifications went undetected, as both the PETN and the non-metallic detonator lacked components that would trigger standard security alarms. This allowed the devices to bypass scrutiny and enter operational use unnoticed.
    \item \textit{Enclosure and Encapsulation:} The entire modified assembly was housed within a black plastic sleeve and enclosed in a metal casing approximately the size of a matchbox. This packaging concealed the PETN sheet and the detonator strip and maintained the appearance of a typical battery. The design effectively masked the modifications and enabled the device to evade visual inspections and blend seamlessly with standard electronic components.
    
\end{itemize}

The walkie-talkie explosions were significantly more deadly than those caused by the pagers due to the larger size and greater internal capacity of the devices. Being bulkier than pagers, walkie-talkies housed more batteries, which provided additional space for integrating a larger quantity of explosive material~\cite{bbc_2}. With more batteries, the walkie-talkies could accommodate a proportionally larger sheet of PETN, substantially increasing the explosive force compared to the smaller, more compact pager. Additionally, the larger size of the device meant that the detonator and the explosives could be arranged more effectively, further enhancing the overall impact of the explosion.

\subsubsection{Remote Triggering through Communication Channels}
\label{subsec_trigger}

With the explosives in place, a remote triggering mechanism was required to activate the devices. The modified pagers and walkie-talkies likely utilized their inherent communication prtocols to facilitate remote detonation.

\begin{figure}[htbp]
\centerline{\includegraphics[width=\columnwidth]{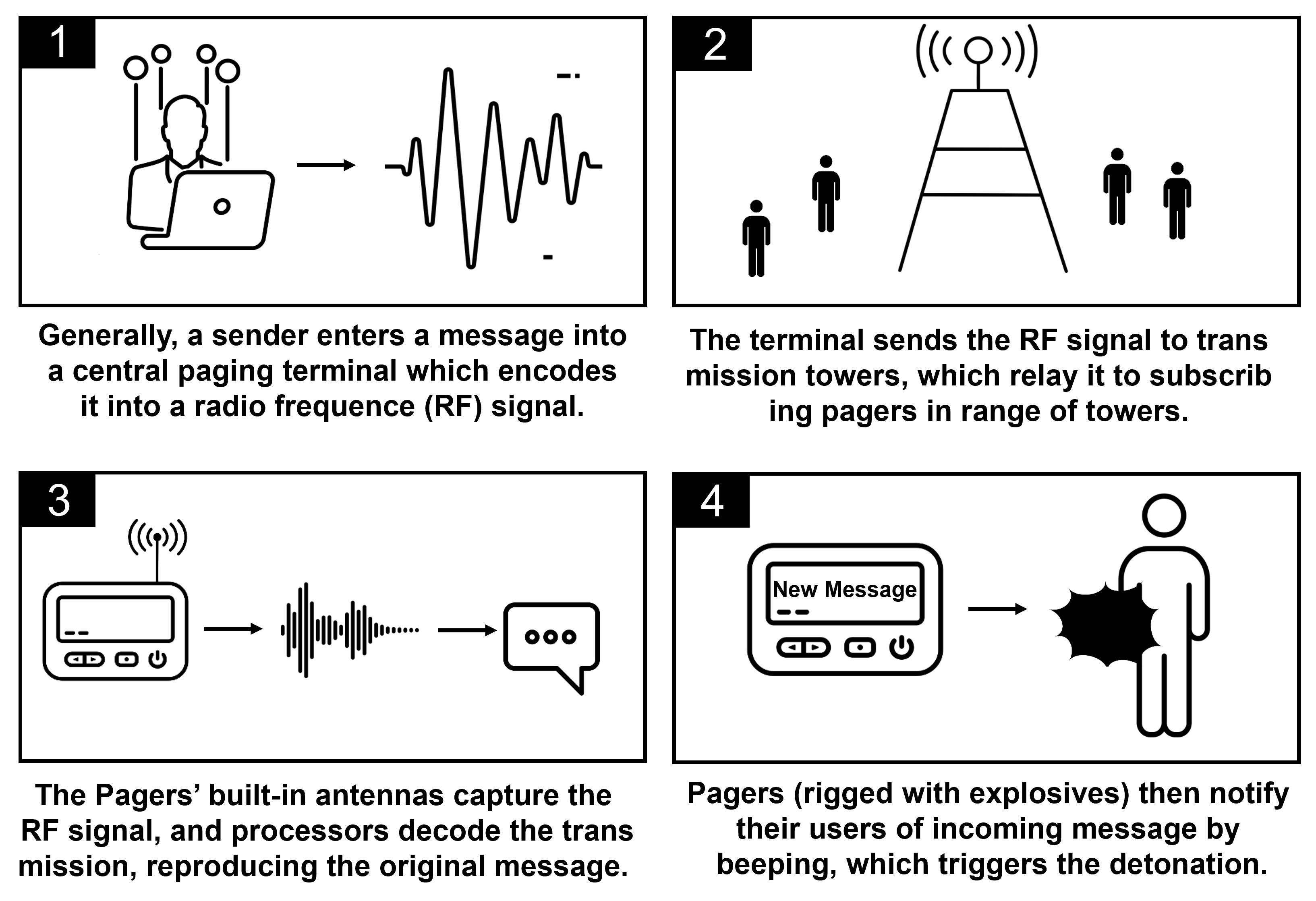}}
\caption{Remote triggering mechanism for pager by using the pager network~\cite{vao}}
\label{fig_trigger}
\end{figure}

\textit{Remote Triggering Mechanism for pagers:} The rigged pagers were integrated into a wireless paging system that uses radio frequency (RF) technology to transmit coded signals from a central controller to receivers. This system enables wide-area communication by relaying encoded signals through a network of transmission towers. The process begins when a sender inputs an alphanumeric message into a central paging terminal, also known as a transmitter controller or paging initiator device. The message is encoded into an RF signal and transmitted via towers to pagers within a specified geographic range, as described in fig.~\ref{fig_trigger}. When the pagers are powered on, their antennas capture the RF signal and send it to internal receivers, which decode the message using embedded processors~\cite{vao}. %If the pagers are programmed to receive the specific message, they alert the users by vibrating or beeping and display the message on their screens.

In the attack, %the perpetrators reprogrammed the pagers to act as detonators. 
upon receiving a specific coded signal, these rigged pagers created a spark in the battery, triggering the detonators, and as a result, the embedded explosives were charged. Unmodified pagers, however, remained unaffected when receiving the same signal. The controller device used to send the detonation signal was likely registered within Lebanon’s wireless paging system, allowing the attackers to blend into the commercial infrastructure and avoid detection~\cite{crypto_m_1}. By operating within the system’s normal parameters, the perpetrators could leverage the network’s transmission towers for signal delivery.%, reducing the risk of exposing their operation.
 Experts, including Bryson Bort, have noted that using a controller device registered in Israel or transmitting signals from Israeli towers would be less likely, as it would require immense power to reach all targeted pagers in Lebanon and would be easily traceable by signal-monitoring systems~\cite{vao}.

\textit{Remote Triggering for Walkie-Talkies:} By leveraging technologies such as Continuous Tone-Coded Squelch System (CTCSS) and Digital Tone-Coded Squelch System (DTCS), the devices could be configured to selectively open their receivers upon detecting specific signals~\cite{bandyopadhyay2010wireless}. Additionally, the integration of Dual-Tone Multi-Frequency (DTMF) coders/decoders enabled precise remote activation using unique tones or codes. This setup allowed adversaries to exploit the devices' existing communication capabilities for covert triggering. A remote signal carrying a distinct tone or coded sequence could have activated an internal circuit to ignite the detonator and set off the explosive charge. By embedding this functionality, the attackers ensured precise control over the timing of the explosions while maintaining the devices' appearance as standard communication tools.
%\section{Our perspective of the attack/vulnerabilities} \label{sec:our_perspective}
\section{Understanding the Vulnerabilities from A Hardware security perspective}%\textcolor{red}{Pantha}}

%\subsection{Various vulnerabilities concluded from the theories}

The pager and walkie-talkie attacks exemplify how multiple hardware security vulnerabilities and attack methods can converge to produce a highly sophisticated operation.  In this case, common communication devices were transformed into instruments of a lethal assault. We will explore how numerous hardware vulnerabilities were exploited.

\subsection{Supply Chain Vulnerabilities}

PCB supply chain vulnerabilities were exploited to create the pager and walkie-talkie attacks.PCB supply chain vulnerabilities, particularly those involving counterfeiting, pose significant risks to hardware security~\cite{asadizanjani2017pcb, asadizanjani2015non}. Unauthorized or counterfeit PCBs—ranging from cloned designs, defective boards, or boards with compromised materials—often enter the supply chain through unverified vendors or intermediaries~\cite{harrison2021malicious}. These counterfeit PCBs can be tampered with during production or assembly, enabling adversaries to introduce malicious modifications, such as embedded hardware Trojans, or compromised signal integrity~\cite{lu2020fics, mehta2022fics}. The global nature of the PCB supply chain—spanning multiple regions and manufacturers with varying security practices—further amplifies these risks by making it challenging to verify the origin and authenticity of PCBs. Such vulnerabilities threaten the reliability and security of hardware systems used in critical sectors such as defense, healthcare, and telecommunications. For example, counterfeit PCBs can degrade system performance, cause premature failures, or serve as entry points for cyberattacks~\cite{true2021geometry, dizon2023framework}. The complexity of the supply chain also allows counterfeit PCBs to bypass quality assurance processes and infiltrate high-security applications~\cite{asadizanjani2021physical}.

\begin{figure}[htbp]
\centerline{\includegraphics[width=\columnwidth]{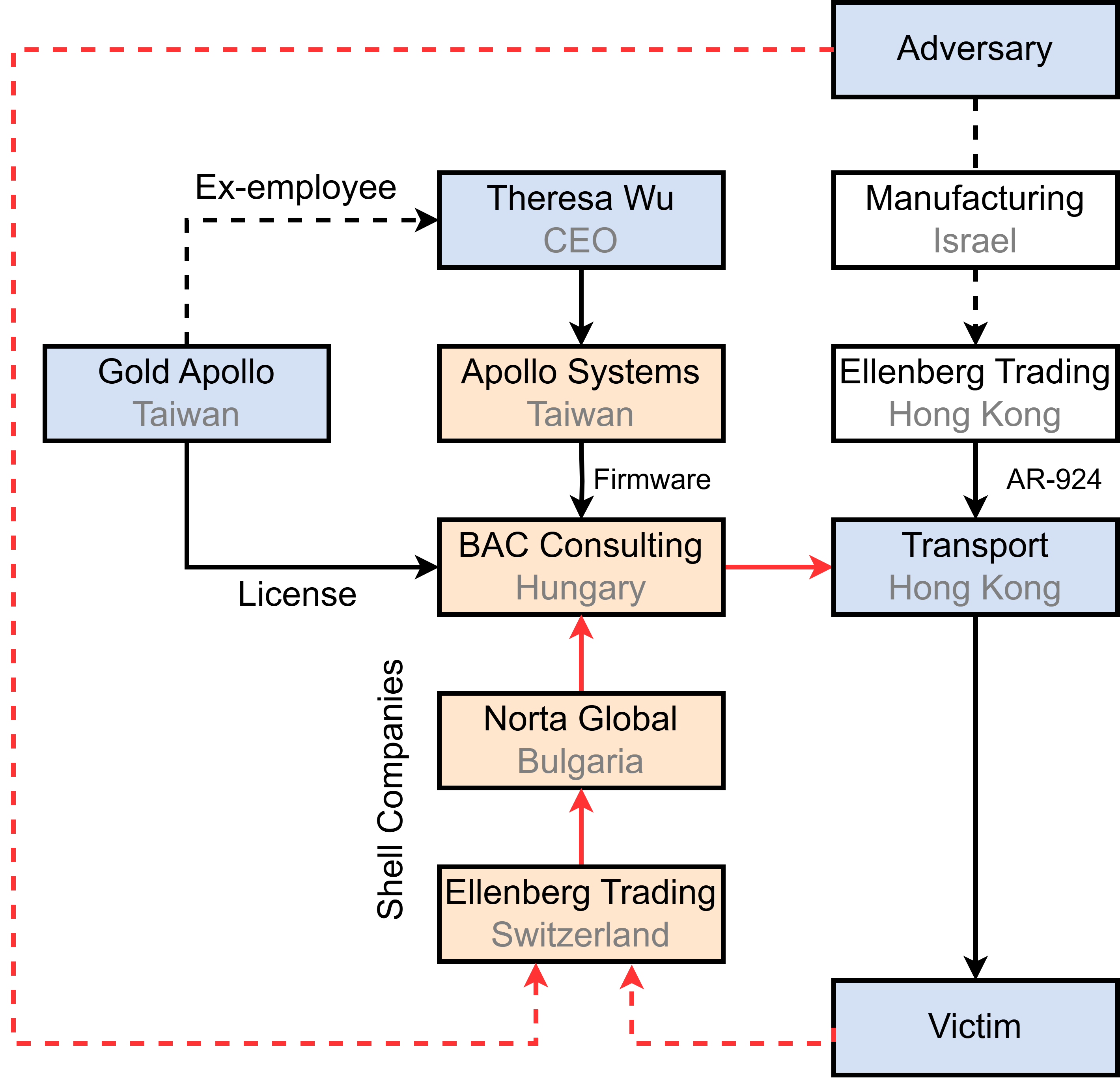}}
\caption{Attackers created a falsified deep supply chain~\cite{crypto_m_1}}
\label{fig_supply}
\end{figure}

Let's consider the pager distribution as a test case to show how PCB supply-chain vulnerabilities were exploited. The attackers constructed a complex and deeply layered false supply chain, as illustrated in Fig.~\ref{fig_supply}. Here, the attackers employed shell companies like Norta Global in Bulgaria and BAC Consulting in Hungary to create a facade of legitimacy. Norta Global, established just weeks before BAC Consulting, facilitated financial transactions and licensing arrangements~\cite{crypto_m_1}. It acted as a key intermediary and handled the distribution of the modified devices. These financial transfers obscured the origin of the funds and shielded the operation from direct scrutiny. Legitimate companies, such as Gold Apollo in Taiwan, were also unwittingly involved. Gold Apollo, a trusted manufacturer of the AR-924 pager, licensed its design and branding to BAC Consulting. %for \$15 per unit. 
This licensing agreement allowed BAC to distribute the devices under the trusted "Gold Apollo" brand, lending credibility to the operation. Gold Apollo’s lack of involvement in manufacturing the tampered devices highlights how licensing agreements can be exploited to legitimize malicious activities.

With no physical manufacturing capabilities, BAC Consulting acted as a financial and logistical intermediary. It subcontracted the production and marketing of the devices to Apollo Systems Ltd., another newly created company in Taiwan. Despite using the Gold Apollo branding, Apollo Systems was reportedly a front with no direct ties to the original manufacturer. The actual production likely took place in a separate, undisclosed location, further complicating the supply chain and concealing the operation’s true origins. The attackers’ use of falsified identities and front companies added another layer of obfuscation. Shell entities like Ellenberg Trading, a Swiss company with no verifiable existence, funneled funds through Norta Global and BAC Consulting. 

This deliberate creation of a fake supply chain showcases how adversaries can exploit the globalized nature of electronics manufacturing. By blending legitimate companies with shell entities, they masked their activities and facilitated the distribution of tampered devices under a trusted brand name.

\subsection{Tampering}

The attackers modified devices at various points in the supply chain, embedding malicious components like explosives and remote-triggering mechanisms~\cite{xiao2014novel}. These changes were carefully integrated to retain the devices' outward appearance and partial functionality, enabling them to bypass routine inspections undetected. By leveraging the devices' existing communication features, such as DTMF codes for remote triggering, the attackers avoided the need for noticeable external modifications.

Reverse engineering played a central role in enabling these modifications~\cite{quadir2016survey}. By deconstructing the devices, the attackers gained a detailed understanding of their internal architecture. This allowed them to identify non-critical components that could be altered or replaced without affecting the devices' primary functionality. The insights from reverse engineering facilitated the introduction of Trojan-like capabilities while minimizing the risk of detection during testing or operation.

This case underscores the significant risks posed by hardware tampering and Trojan insertion. The lack of tamper-evident designs, secure manufacturing protocols, and traceability in the supply chain provides adversaries with multiple opportunities to introduce malicious modifications. %Fig.~\ref{fig_tampering} highlights the critical stages where tampering is most likely to occur. 
To address these vulnerabilities, the adoption of tamper-resistant packaging, security aware design principles, and rigorous supply chain testing is essential~\cite{contreras2013secure, xiao2014novel}. Without these measures, critical systems remain exposed to covert modifications that threaten both their functionality and security.

\section{Extension of the battery-based attacks} %Shang
\label{sec:extension}
%Our POV of where these attacks can be extended

In addition to the physical and supply chain attacks discussed above, the extended cyber attack surface, which is created by the use of battery management systems(BMS) for more comprehensive and accurate management of the battery pack, has become more of a concern. This section will present our perspectives on how these attacks on pages and walkie-talkies can be extended to the cyber-physical interface of the BMS.

\subsection{Architecture of Battery and Battery Managment Systems Based Devices }
\begin{figure}[t]
    \centering
    \includegraphics[width=\columnwidth]{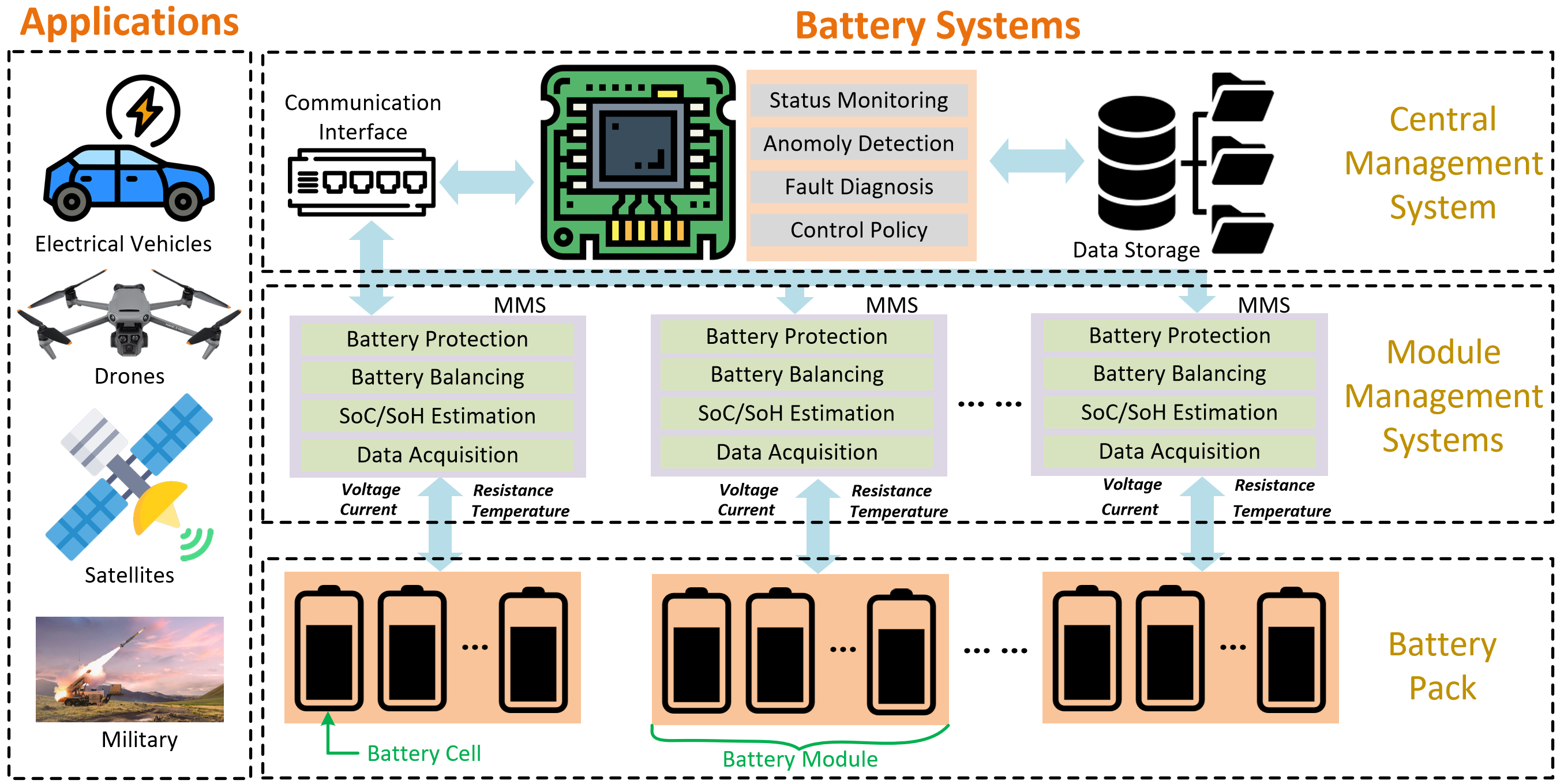}
    \caption{Overview of a general smart battery system.\cite{cryptoeprint:2024/211}}
    \label{fig:BMSoverview}
\end{figure}

Battery systems are widely used across various applications, including mission-critical equipment, Electric Vehicles (EVs), and personal smart devices. They have multiple types and structures, each with specific features to accommodate the unique demands of particular products. Fig. \ref{fig:BMSoverview} shows that a smart battery system typically comprises three main components: a central management system, module management systems, and battery packs, 

\subsubsection{Battery Pack} As the power source of the applications, Li-on battery packs differ in parameters including size, voltage, and capacity. For instance, devices with intensive energy demands such as drones and electrical vehicles have higher demands on the lifetime and the density of power of the battery pack. In contrast, smaller devices like personal electronics emphasize the most on the battery pack being lightweight, which means storing a considerable amount of energy in a relatively compact package. A battery cell serves as the fundamental unit of a battery pack, converting chemical energy into electrical power. Multiple cells are mechanically connected in series or parallel to form a battery module, which functions as a serviceable unit. These modules, along with optional peripheral components such as temperature control systems, collectively make up the battery pack.

\subsubsection{Module Management System}
Since the battery pack is unlikely to function independently in an advanced system, it requires continuous monitoring and real-time adjustments to ensure system safety during operation.
As a result, module management systems (MMS) including microelectronic-based control units and sensors are utilized for local battery management., as demonstrated in Fig. \ref{fig:BMSoverview}. The presence of MMS instances allows for fine-grained management of each battery module within the pack and even individual battery cells within the modules. The primary functions of MMSs are illustrated as follows:

\begin{itemize}
    \item \textit{Data Acquisition:} This enables the quantification and collection of status statistics for the target battery module and each individual battery cell within it. 
    \item \textit{SoC/SoH Estimation:} It helps estimate the percentage of remaining energy relative to the total battery capacity, as well as the remaining capability and level of degradation of an in-use battery, commonly referred to as the state-of-charge (SoC) and state-of-health (SoH).
    \item \textit{Battery Protection:} For mission-critical infrastructures, battery safety is of utmost importance. The MMS can detect anomalies such as over-voltage or over-temperature, which could lead to issues like pack leakage or even explosions. By analyzing battery status, the MMS can activate safeguards, such as cutting off the MOSFET switch, to disconnect the battery from the payload and prevent further consequences.
    \item \textit{Battery Balancing:} Additionally, the collected statistics enable effective planning for battery balancing, which can extend the lifespan of the entire pack. Due to inherent variations and differing aging patterns among individual cells—caused by manufacturing uncertainties—the overall capacity of the pack can be maximized by balancing each cell to achieve an equivalent SoC.
\end{itemize}

\subsubsection{Central Management System} Other than the functions provided by the local MMS, there is also a need for more sophisticated algorithms for advanced battery management. As a result, there is usually a central management system (CMU) connected to the MMS to enable the joint establishment of the battery management system. 
Depending on the master system’s physical layer, either a serial interface or a wireless channel can be used. But the performance of a standard microprocessor is limited to gigabits per second, and both storage and scalability face constraints due to the high number of battery cells in advanced applications like stationary charging stations or electric vehicles \cite{naseri2023cyber}. On the other hand, state-of-the-art data-driven battery management algorithms, such as machine learning (ML)-based solutions, rely heavily on large volumes of historical data to understand and predict battery behaviors. These algorithms require substantial computational power and memory allocation to process and analyze these data effectively\cite{lee2020deep}. To deal with such challenges, more investments are shifting the architecture of the battery system from local devices to cloud BMS \cite{BOSCHBMS}. 
A digital twin-based virtual central BMS is typically hosted on high-performance cloud servers\cite{li2020digital}, while MMS are implemented on lightweight Internet-of-Things (IoT) devices for edge processing. The battery data is transmitted to the cloud platform via wireless protocols. To improve efficiency and adaptability, resource-intensive tasks like data storage, statistical processing, and machine learning model deployment are offloaded to the cloud management unit (CMU).\cite{li2020digital}. 
However, as hybrid cyber-physical systems (CPS), battery systems are vulnerable to various security threats and concerns.

\subsection{Taxonomy of Attacks on Battery and BMS Based Systems}
% \subsection{Taxonomy of possible attacks}

\begin{figure}[t]
    \centering
    \includegraphics[width=\columnwidth]{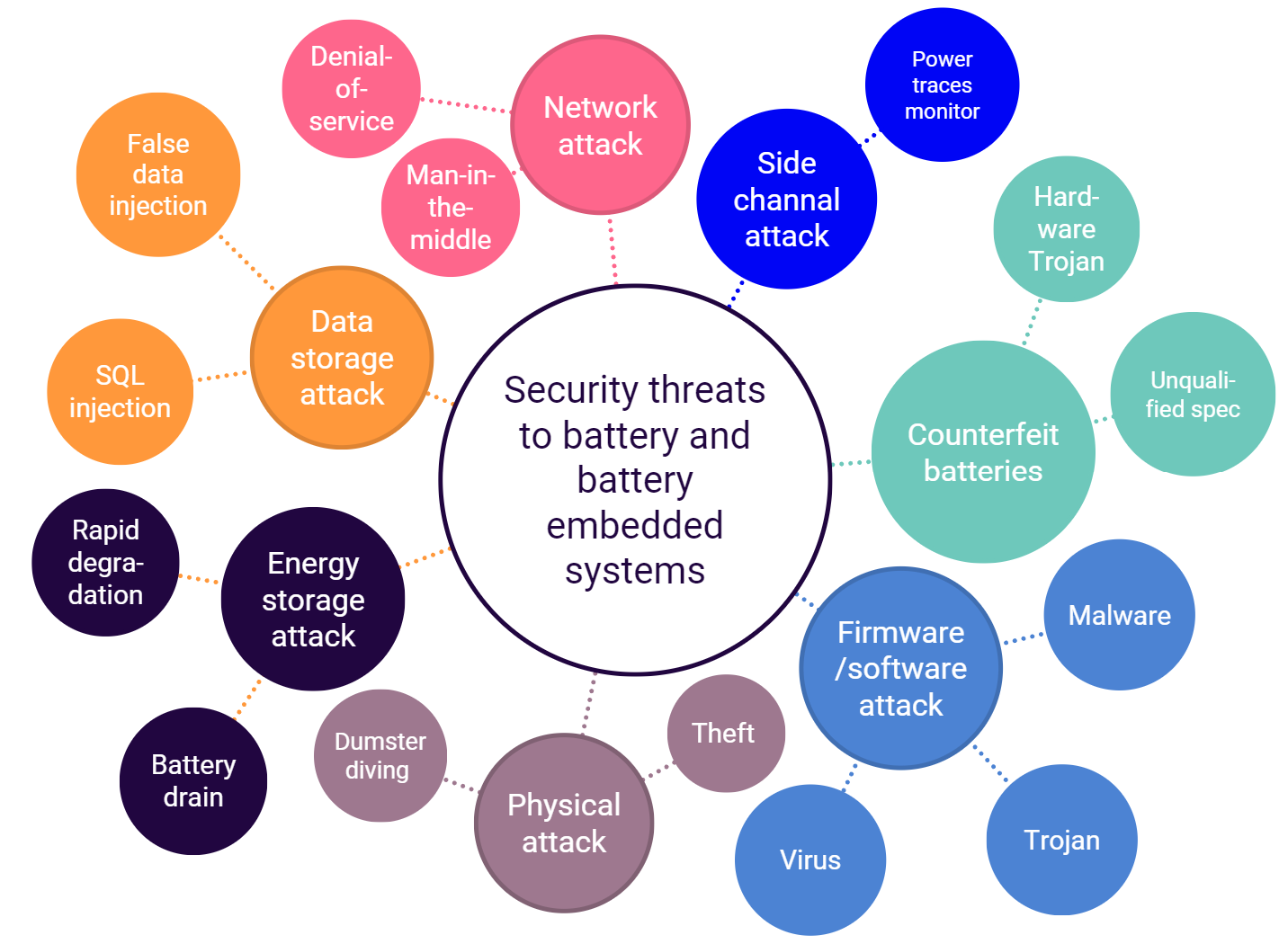}
    \caption{Taxonomy of security threats to battery and battery embedded systems. \cite{KHARLAMOVA2023107795, Spyingbatteries, 8964396, 10263843}}
    \label{fig:SecurityThreats}
\end{figure}

Fig. \ref{fig:SecurityThreats} depicts the taxonomy of security threats to battery and battery-embedded systems, as detailed as follows.

\begin{enumerate}
    \item  \textit{Physical Attack:} This type of attack targets the physical access of the battery system by either theft or dumpster diving, which could lead to various problems like information leakage and reverse engineering \cite{8964396}. Nowadays, the BMS is usually specifically designed and customized to meet the unique requirements of individual applications \cite{EPECReport}. Reverse engineering of such customized BMS designs can allow adversaries to gain full access to the system's architecture, exposing security vulnerabilities and potentially leading to the theft of the BMS intellectual IP design \cite{TeslaBMSRE}. Moreover, physical fault injection attacks pose significant risks. For instance, an external electromagnetic (EM) source can induce large Eddy currents in the on-chip power distribution network, disrupting the BMS circuitry by causing setup/hold time violations in a contactless manner. 

    \item  \textit{Firmware/Software Attack:} This type of attack targets the firmware or software of the battery-embedded system. Attackers may use tactics like phishing emails, social engineering, or infected USB drives to install malware, Trojans, or viruses. These malicious actions can lead to data theft, system malfunctions, or even physical damage to the batteries \cite{lopez2017security}.

    \item  \textit{Counterfeit Batteries:} Counterfeit batteries often differ from genuine ones in terms of specifications. They may have reduced capacity, shorter lifespans, or lack critical safety components, making them potentially hazardous. In addition, counterfeit batteries may contain malicious hardware Trojans, which could result in even more severe issues \cite{Blast}.

    \item  \textit{Side-Channel Attack:} Researchers have discovered that a malicious battery can extract various types of information from a device by exploiting side-channel data, such as by continuously monitoring power traces \cite{9401542}. The activity of the battery-powered system can be predicted by analyzing the power traces of a battery \cite{5751489}. For example, monitoring the power consumption of a cell phone can reveal details about its usage patterns and the specific types of activities being performed \cite{7782756}.

    \item  \textit{Network Attack:} Network attacks usually target the communication channels between the battery system and the user or between the battery and the controller. An attacker may overload the physical or network connections to execute a Denial-of-Service (DoS) attack, making the battery or the entire system inaccessible. Alternatively, the attacker could intercept or manipulate communication between entities that believe they are directly interacting, breaching stored data or executing unauthorized operations, such as installing malware, Trojans, or viruses \cite{9507536, 1276868}.

    \item  \textit{Data Storage Attack:} A data storage attack targets the unit for data storage of the battery system. The attacker may attempt to alter or breach the stored data using various methods, including deploying malware, Trojans, or viruses, gaining unauthorized physical access, or interfering with the database through techniques like SQL injection attacks \cite{9284586}. 

    \item  \textit{Energy Storage Attack:} Energy storage attacks usually target the energy storage unit of the battery system. The attacker may attempt to deplete the battery's energy by flooding it with fake packets, forcing the system to continually send acknowledgment (ACK) signals. Alternatively, the attacker could inject false data into the battery's state estimation process, causing the controller to misinterpret the battery's condition and engage in excessive charging or discharging, which accelerates battery degradation \cite{RUAN2022100158}.

    \end{enumerate}

\section{Existing Physical Inspection Techniques and their challenges 
}

Physical inspection technologies are indispensable in modern security frameworks, especially in environments prone to threats such as terrorism or smuggling. These technologies have evolved to detect concealed weapons, explosives, and unauthorized electronics. However, the sophistication of modern threats, including devices such as modified pagers and walkie-talkies, demands heightened inspection capabilities. Such devices can be exploited for covert communication or as enclosures for improvised explosive devices (IEDs). Consequently, while the existing technologies form the backbone of security operations, their limitations become apparent when facing threats involving innocuous-looking, widely-used electronics.

\subsection{X-Ray Imaging}

X-ray imaging has long been a cornerstone of security inspection systems, primarily due to its ability to rapidly and non-invasively generate images of an object’s internal structure ~\cite{asadizanjani2017pcb, royapplications}. The technique relies on the principles of electromagnetic radiation, specifically X-rays, which have wavelengths in the range of 0.01 to 10 nanometers. When X-rays pass through an object, they are absorbed at different rates depending on the material's density and atomic number (Z). High-density materials, such as metals, strongly attenuate X-rays, appearing as bright regions on the detector, while low-density materials like plastics and organic compounds allow more X-rays to pass through, appearing darker.This fundamental reliance on material density and atomic number is both a strength and a limitation.Some of the examples of high-resolution x-ray imaging used to detect the explosive materials are shown in Figure 9.  While it excels at identifying dense metallic objects, conventional X-ray systems struggle with detecting threats embedded in non-metallic or low-density materials ~\cite{ahi2015terahertz}. This limitation played a critical role in the failure to detect explosive devices hidden within pagers and walkie-talkies during the attacks in Lebanon.

\begin{figure}[htbp]
    \centering
    \includegraphics[width=\columnwidth]{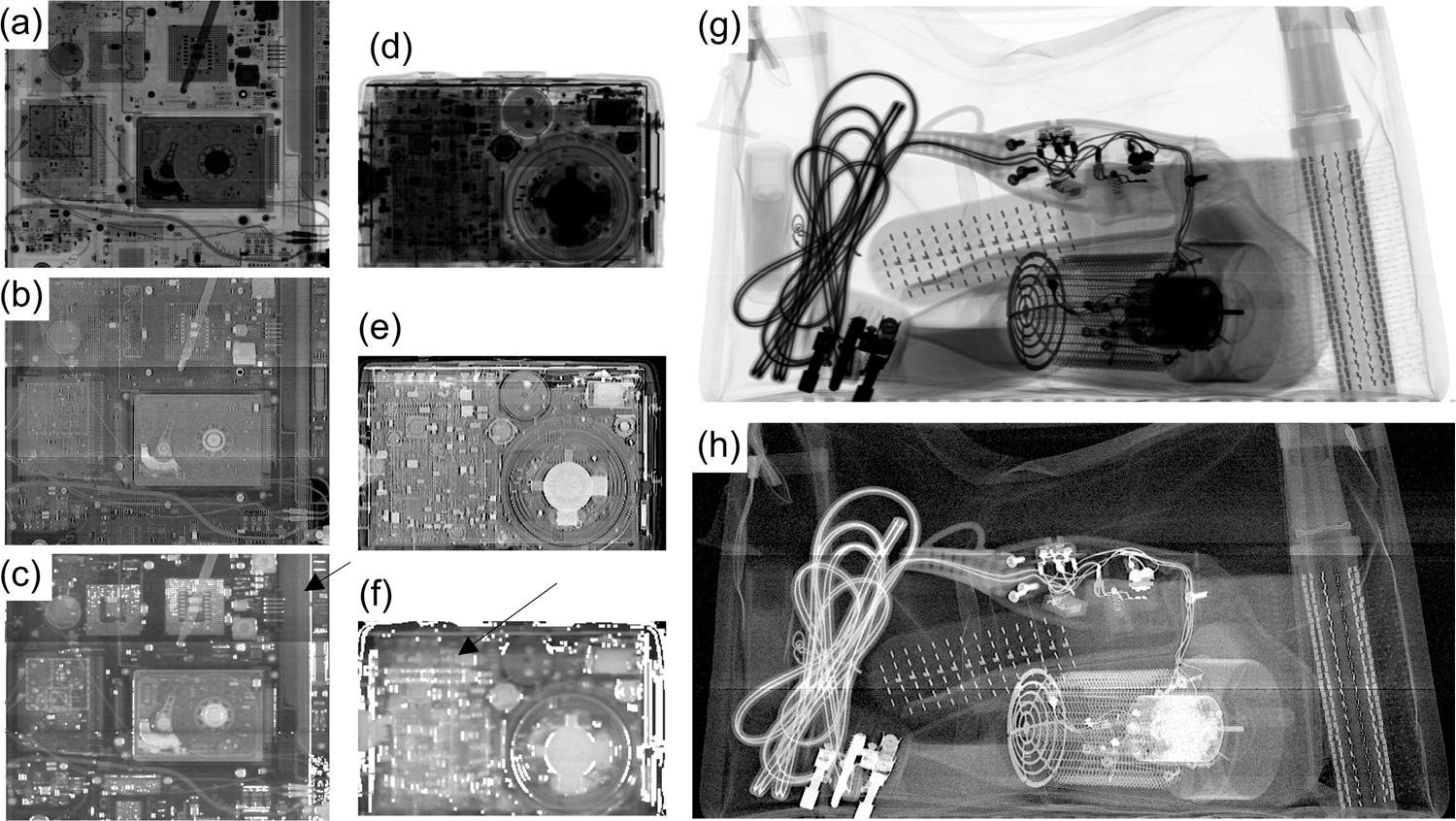}
    \caption{Attenuation (a, d), dark-field (b, e) and ratio (c, f) images of the laptop and mobile phone (respectively) in which a small quantity of C4 was concealed are shown\cite{partridge2022enhanced}}
    \label{x-ray}
\end{figure}

X-ray detection relies on two primary interactions: the photoelectric effect and Compton scattering ~\cite{cooper2004x}. The photoelectric effect dominates in high-Z materials like metals, where X-rays are absorbed, creating high-contrast images. Conversely, Compton scattering prevails in low-Z materials, reducing attenuation and contrast. Explosives and plastic casings, being low-Z materials, often fail to produce distinct signals, effectively camouflaging threats ~\cite{runkle2009photon}. Additionally, X-rays of varying energies interact differently with materials; low-energy X-rays enhance detection of low-Z substances but lack penetration for thick objects, while high-energy X-rays penetrate deeply but are less effective in distinguishing low-Z materials. Conventional systems typically operate at a fixed energy level, balancing penetration and contrast, which limits detection capabilities for concealed explosives. Advanced dual-energy or material-specific imaging systems, which analyze absorption at different energy levels, can improve detection but are not widely implemented in standard security setups. This limitation was evident during the Lebanon inspections, where the absence of dual-energy analysis hindered explosive detection. Furthermore, internal components in cluttered devices like pagers and walkie-talkies scatter X-rays, generating noise that obscures small amounts of explosive material, a challenge exacerbated by the intricate designs of modern electronics.

In the Lebanon attacks, these challenges for X-ray imaging in this context were multifaceted:

\subsubsection{Device Composition and Concealment}
Pagers and walkie-talkies consist primarily of low-density materials like plastic casings, polymer-based circuit boards, and lithium-ion batteries. These materials exhibit low attenuation of X-rays, making them appear similar to benign objects in X-ray scans. When explosive materials such as plastic explosives (e.g., PETN or RDX) are integrated into the internal cavity of such devices, their density and X-ray attenuation characteristics closely match those of the surrounding materials. This blending effect makes it exceedingly difficult for X-ray systems to differentiate between benign electronics and a concealed threat.

\subsubsection{Explosive Material Integration}
In the attacks, the explosive materials were strategically integrated into the internal components of the pagers and walkie-talkies, such as replacing the battery or filling void spaces within the circuit housing. Plastic explosives like PETN are known for their low atomic number and high energy density, which make them challenging to detect using density-based X-ray techniques ~\cite{wells2012review}. Unlike metals or traditional explosives, these materials do not produce the stark contrast required for clear identification.

\subsubsection{Limited Material Differentiation}
Conventional X-ray imaging operates primarily on a grayscale spectrum, representing different levels of X-ray absorption. However, this method lacks the specificity to distinguish between materials with similar densities. For example, the polymer casing of a pager may absorb X-rays similarly to certain explosives, rendering the concealed threat virtually invisible in a standard X-ray image.

\subsubsection{Operator Dependency and Cognitive Load}
X-ray inspection heavily relies on trained operators to identify suspicious objects by visually interpreting the images. In high-pressure environments like airports or border checkpoints, where thousands of items are scanned daily, the likelihood of overlooking subtle anomalies increases. In the Lebanon attacks, the attackers exploited this human limitation by designing the modified devices to resemble commercially available products, further reducing the likelihood of detection.

The attackers in Lebanon demonstrated an acute understanding of the limitations of X-ray imaging. By using devices that naturally contained batteries, circuit boards, and other benign components, they ensured that their modifications blended seamlessly into the device’s internal structure. The use of non-metallic explosive materials compounded the issue, as these substances did not produce the sharp contrasts typically associated with traditional threats like metal weapons or metallic IED components.
Moreover, the placement of explosive materials in voids or as replacements for legitimate components created an appearance consistent with normal electronics. This strategic design exploited both the physical limitations of X-ray technology and the cognitive biases of operators, who are trained to identify common threats rather than unconventional designs.

\subsection{Computed Tomography (CT) Scanning}

Computed Tomography (CT) scanning is an advanced imaging technology widely recognized for its ability to generate detailed three-dimensional (3D) images of objects ~\cite{true2021geometry}. Unlike traditional X-ray imaging, which provides a two-dimensional view, CT systems capture multiple cross-sectional images by rotating an X-ray source around the object. These images are then reconstructed computationally to create a volumetric representation, allowing inspectors to analyze the internal structure of objects layer by layer. CT scanning has become increasingly important in security settings for identifying complex threats concealed in luggage, packages, or electronic devices ~\cite{velayudhan2022recent}.

While CT systems offer significant advantages over conventional X-ray imaging, such as better resolution and material discrimination, they are not without limitations. These limitations became apparent in the context of the pager and walkie-talkie attacks in Lebanon, where modified consumer electronics containing explosive materials evaded detection even under sophisticated imaging systems. In the attacks in Lebanon, the use of modified pagers and walkie-talkies to conceal explosive materials presented a unique challenge to CT scanning systems. The inability of CT to detect the threats stemmed from several key factors:

\subsubsection{Integration of Explosives into Benign Electronics}
The attackers exploited the internal structure of consumer electronics by embedding explosives into voids or replacing legitimate components like batteries or capacitors. Materials such as PETN (pentaerythritol tetranitrate) or RDX (Research Department Explosive), commonly used in plastic explosives, have low densities and attenuation properties similar to those of the surrounding polymers or plastics. In CT images, this similarity results in a lack of contrast, making it difficult to distinguish the explosives from benign components ~\cite{harding2004x}.

\subsubsection{Homogeneity in Material Properties}
CT scanning excels at identifying differences in material density and composition. However, in the case of the Lebanon attacks, the materials used for the explosives and the electronic housing exhibited nearly identical attenuation coefficients. For example, the plastic casing of a pager or walkie-talkie, combined with the low-density nature of the explosives, produced uniform attenuation signals that did not raise alarms. This homogeneity effectively masked the presence of the explosive material.

\subsubsection{Complexity of Internal Structures}
Modern electronics, including pagers and walkie-talkies, are designed with intricate internal layouts featuring circuit boards, wires, batteries, and other components. In a CT scan, this complexity generates a high degree of structural noise, making it difficult to identify anomalies ~\cite{withers2021x}. The compact and dense arrangement of legitimate electronic components can overshadow the subtle presence of additional materials, such as explosives.

\subsubsection{Lack of Material-Specific Differentiation}
While dual-energy CT systems can provide better material discrimination, their resolution is often insufficient for identifying low-Z materials like plastic explosives when embedded within complex electronics. The atomic numbers of PETN or RDX are close to those of common polymers, meaning the system cannot reliably differentiate between them. This limitation was likely exploited by the attackers to ensure the concealed explosives blended seamlessly with the device's internal structure.

\subsubsection{Operator Dependency and Training Gaps}
Although CT systems generate 3D images, their interpretation still relies on human operators. Recognizing a concealed explosive device within a benign-looking pager or walkie-talkie requires extensive expertise and familiarity with unconventional threats. In high-pressure scenarios, such as airport or border inspections, operators may focus on more common threats, inadvertently overlooking subtle discrepancies in complex electronic devices.

By embedding explosives in components that naturally align with the device's original design, the attackers ensured the modifications did not create visible anomalies in CT scans. Furthermore, the use of low-density explosives and careful placement within the device's structure exploited the inherent weaknesses of CT systems in differentiating materials with similar attenuation properties.

\subsection{Explosives Trace Detection (ETD)}

Explosives Trace Detection (ETD) is a critical technology used to identify trace amounts of explosive materials through chemical analysis. The process typically involves collecting particles or vapors from an object’s surface, followed by analysis using techniques like Ion Mobility Spectrometry (IMS) or Gas Chromatography-Mass Spectrometry (GC-MS). IMS, the most common method, identifies explosives by measuring the ion mobility patterns of molecules subjected to an electric field, while GC-MS offers higher specificity by separating and identifying chemical compounds based on their structure and mass. ETD is widely deployed at airports, border checkpoints, and other security settings, where its ability to detect even minute traces of explosive residue makes it an indispensable complement to imaging systems. However, its effectiveness relies on certain physical and environmental factors, including the presence of detectable residues on the surface or vapors emitted from the explosive material.

In the Lebanon attacks, where explosives were hidden within pagers and walkie-talkies, ETD systems faced significant challenges that rendered them ineffective. The explosive materials, such as PETN or RDX, were sealed inside the devices, preventing the release of detectable particles or vapors to the exterior. With low vapor pressures, these explosives emitted minimal detectable molecules, especially under standard inspection conditions, further complicating detection. Additionally, the internal structure of the electronics likely masked or interfered with any trace signals, and meticulous handling by the attackers ensured no residue was left on external surfaces. The encapsulation of explosives and their low volatility, combined with the inherent limitations of sampling only external surfaces, meant that the ETD systems failed to identify the concealed threats. Overcoming these limitations will require advancements such as deeper sampling techniques, enhanced sensor sensitivity, and integration with imaging technologies to detect concealed explosive threats effectively.

\label{sec:exist_countermeasures}
\section{Emerging Physical Inspection Techniques }

Emerging physical inspection techniques, such as terahertz (THz) imaging ~\cite{guerboukha2018toward,xi2024enhancing}, quantum imaging ~\cite{lugiato2002quantum}, and advanced neutron scanning ~\cite{anderson2009neutron}, hold significant promise in detecting concealed explosives in complex devices like the modified pagers used in the Lebanon attacks. These techniques address limitations in traditional methods like X-ray and CT scanning, which struggle with detecting non-metallic explosives or materials embedded within intricate electronic systems. THz imaging, for example, leverages molecular absorption signatures to identify explosive compounds like PETN and RDX, while quantum imaging utilizes entangled photons to provide high-resolution, non-invasive imaging of internal structures. Neutron-based methods, such as neutron activation analysis, offer enhanced detection capabilities for dense materials like explosives by probing atomic composition. Combined, these advanced techniques can overcome the detection challenges faced by conventional methods, enabling more effective screening and identification of hidden threats in everyday objects, thus improving physical security protocols in high-risk environments. These emerging techniques have been discussed further in detail below:

\begin{figure}[htbp]
    \centering
    \includegraphics[width=\columnwidth]{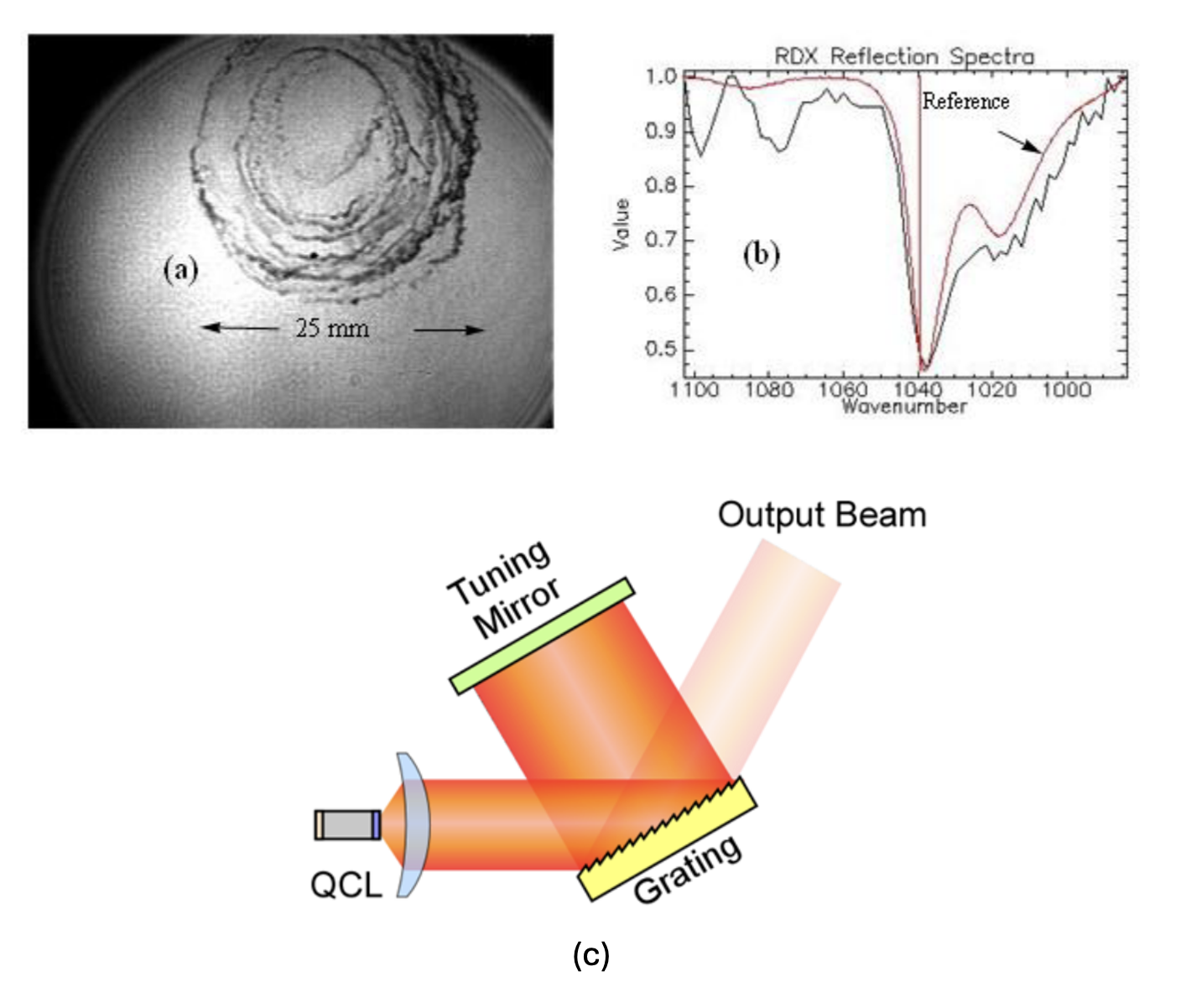}
    \caption {A grayscale image of the RDX residue shown in (a), and an example spectrum of a 3 x 3 patch of pixels in (b). (c) Schematics of tunable external cavity quantum cascade laser illuminator  \cite{bernacki2010standoff}}
    \label{quantum}
\end{figure}

\subsubsection{Quantum Imaging}

Quantum imaging offers a revolutionary solution to address the limitations of traditional inspection techniques for detecting concealed threats, such as the explosives used in the pager and walkie-talkie attacks in Lebanon. Utilizing principles like quantum entanglement and ghost imaging, this approach surpasses the constraints of X-rays and other conventional systems ~\cite{gilaberte2019perspectives}. Quantum imaging employs entangled photon pairs, where one photon interacts with the object and the other is analyzed separately, enabling high-resolution imaging with reduced noise and greater material specificity. This allows for the non-invasive inspection of electronic devices, even through opaque or layered structures, without requiring direct interaction or high-energy radiation. Previous works have shown the capabilities of using quantum lasers for the explosives materials detection as shown in Figure 10.  Unlike X-ray or CT imaging, which primarily detect density variations, quantum imaging can capture subtle differences in molecular or structural properties, making it capable of identifying concealed explosives with greater confidence.

One of the primary advantages of quantum imaging is its ability to overcome the penetration and specificity challenges faced by other methods. For instance, ghost imaging excels in environments where other modalities fail due to noise or limited penetration depth, such as scenarios involving metallic shielding within electronic devices. Additionally, quantum techniques are inherently less invasive, as they do not rely on ionizing radiation or require surface residue for detection, unlike mass spectrometry or X-ray backscatter. By leveraging quantum-enhanced sensitivity to molecular differences and exploiting phenomena like quantum superposition and interference, quantum imaging can detect minute structural irregularities or identify chemical compositions indicative of explosives. As this technology matures, its integration into security frameworks could offer unparalleled precision and reliability in detecting concealed threats, setting a new standard for physical assurance against sophisticated attacks.

While quantum imaging holds significant promise, it is still an emerging field, with practical deployment in security scanning likely several years away. Currently, most quantum imaging techniques are used in controlled laboratory environments. However, as technology advances, quantum imaging could become a powerful tool for detecting concealed explosives and modifications in complex electronic devices, offering highly detailed, non-invasive imaging that could be critical in counter-terrorism and security fields.

\subsubsection{Neutron Imaging}

Neutron imaging offers a promising solution for detecting hidden explosives in devices like the pagers used in the recent Lebanon attacks, addressing many of the shortcomings of conventional physical inspection methods. Unlike X-ray or terahertz imaging, which rely on the interaction of electromagnetic waves with matter, neutron imaging utilizes the scattering and absorption properties of neutrons, which are sensitive to light elements like hydrogen and nitrogen—key components of explosives like PETN and RDX. Neutron imaging works by directing a beam of neutrons at an object and analyzing how the neutrons scatter or are absorbed by the materials  \cite{anderson2009neutron, kardjilov2011neutron}. Explosive materials, which often contain high amounts of hydrogen, carbon, and oxygen, exhibit distinct scattering patterns that can be differentiated from the background electronic components of devices like pagers or walkie-talkies. This technique can provide detailed internal images of objects, revealing hidden explosive devices even if they are concealed within layers of non-metallic materials or complex circuitry.

The primary advantage of neutron imaging over other inspection techniques is its ability to penetrate dense materials, including metals and plastics, which often impede other forms of radiation like X-rays or terahertz waves. Since neutrons interact with nuclei in a different manner than X-rays, they can detect material properties that are otherwise opaque to electromagnetic waves, making them invaluable for identifying explosives embedded within electronic devices. Furthermore, neutron imaging can be combined with other neutron-based techniques, such as neutron activation analysis, to determine the exact composition of suspicious materials with high precision. This level of sensitivity and accuracy can be crucial in scenarios where conventional X-ray or terahertz methods might fail due to insufficient penetration depth or inability to differentiate between benign and hazardous materials.

\begin{figure}[htbp]
    \centering
    \includegraphics[width=\columnwidth]{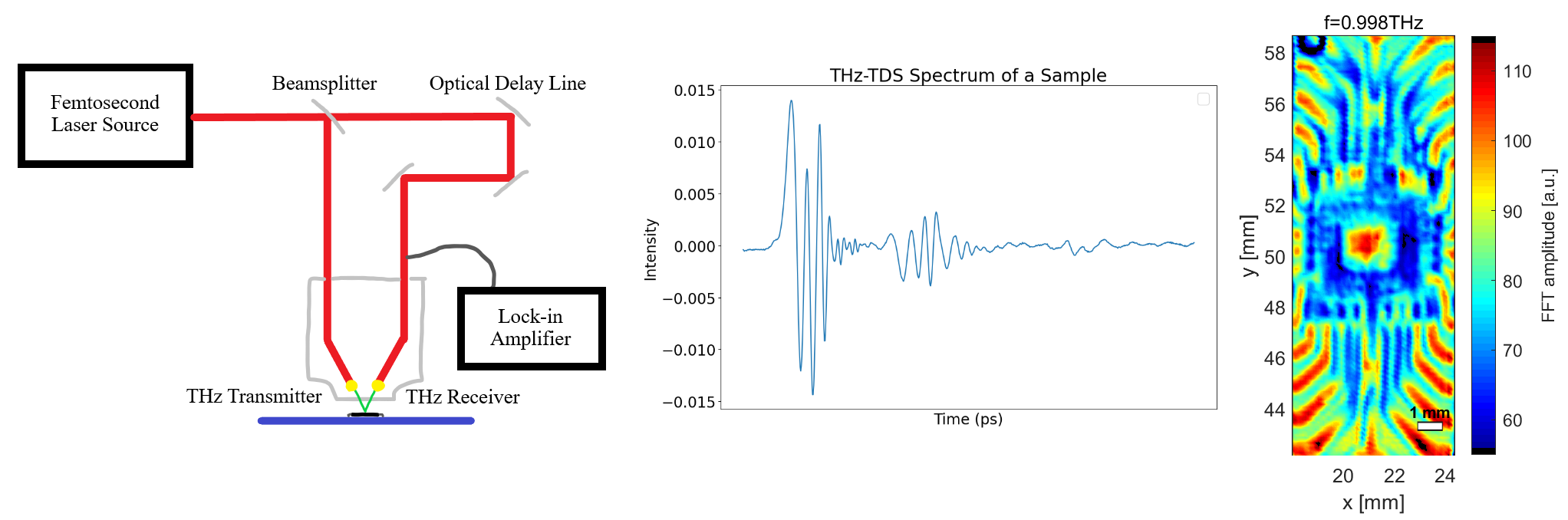}
    \caption{Diagram of near-field THz-TDS tool (Left). Example of spectral data (Middle). Example 2D reconstruction of spectral data (Right)\cite{craig2024terahertz}}
    \label{tetrahertz}
\end{figure}

While neutron imaging shows significant potential for detecting concealed explosives, its widespread adoption for real-time security screening faces several challenges. Neutron sources, which often involve nuclear reactors or particle accelerators, are large, expensive, and not easily portable. This limits the feasibility of neutron imaging in fast-paced, high-throughput environments like airports or public transportation hubs. Furthermore, the need for specialized detectors and high radiation shielding presents logistical and safety concerns. However, research into compact neutron sources and improved detection technology is ongoing, with the aim of developing portable systems that can be deployed in real-world scenarios. While significant work remains to miniaturize and optimize these systems for practical use, neutron imaging could eventually become a critical tool in the fight against terrorism, complementing other inspection technologies and providing a more reliable method for detecting hidden threats in everyday objects.

\subsubsection{Terahertz (THz) imaging}

Terahertz (THz) imaging is a promising technique for detecting concealed explosives in devices like the pagers used in the recent Lebanon attacks, addressing many of the challenges faced by conventional inspection methods such as X-ray or CT imaging. Terahertz radiation lies between the infrared and microwave regions of the electromagnetic spectrum and exhibits the unique ability to penetrate non-metallic materials, such as plastics, clothing, and even some composites, while reflecting off metals as shown in Fig. 11. This property makes THz imaging particularly effective for inspecting objects like pagers or walkie-talkies, which often contain complex, multi-layered, and non-metallic components. The key to its success lies in the molecular absorption signatures of materials, such as explosives like PETN and RDX, which exhibit distinct absorption peaks in the THz frequency range. These signatures are unique to certain chemical compounds, allowing for the identification of explosives even when concealed within other materials, making THz imaging a viable method for non-invasive detection.
In real-world scenarios, THz imaging could be deployed at security checkpoints, such as airports or public transport hubs, to screen everyday electronic devices for hidden threats. The technique's ability to provide high-resolution imaging without the ionizing radiation associated with X-rays makes it an attractive alternative for security purposes. However, there are challenges to overcome before THz imaging can be used in real-time, large-scale security applications. Current THz systems often require bulky, expensive equipment that limits their portability and deployment in high-throughput environments. Additionally, environmental factors such as humidity can affect the performance of THz waves, as water vapor absorbs THz  \cite{xi2024thz}. To make THz imaging more practical, research is focused on developing compact, robust, and cost-effective systems, as well as improving their sensitivity and calibration to environmental conditions. With advancements in these areas, THz imaging could become a key tool in the detection of hidden explosives, providing an effective complement to existing security technologies.
\label{sec:emerg_countermeasures} 

\section{Research Opportunities} 
\label{sec:roadmap}

The global electronics supply chain faces unprecedented challenges in ensuring security and trustworthiness, particularly as adversaries exploit its inherent vulnerabilities. Recent pager and walkie-talkie attacks in Lebanon have underscored the sophistication and lethality of hardware-based threats. These incidents exemplify how even everyday electronic devices can be weaponized, revealing critical gaps in existing frameworks. This section explores key research challenges and highlights emerging approaches to protect the electronics supply chain and improve tampered device detection.

\subsection{Challenges in Protecting the Electronics Supply Chain}

\begin{enumerate}
    \item \textit{Complexity and Globalization of Supply Chains:} The electronics supply chain spans multiple countries, with components sourced, manufactured, and assembled in diverse regions. This complexity creates opportunities for unauthorized tampering or the insertion of counterfeit parts at various stages, from design to distribution. Licensing agreements, shell companies, and unverified intermediaries exacerbate the difficulty of tracing and securing the origins of components, as evidenced by the exploitation of Gold Apollo’s licensing agreement in the pager attacks. Ensuring end-to-end traceability in such a fragmented ecosystem remains a formidable challenge.
    \item \textit{Tamper-Resistant Design and Detection Limitations: }Existing electronic designs often lack robust tamper-evident features, enabling attackers to modify devices while maintaining their outward functionality and appearance. For instance, embedding explosives or remote-triggering mechanisms into legitimate components highlights the inadequacy of current physical inspection protocols. The absence of tamper-resistant packaging and secure manufacturing practices amplifies the risk of covert modifications.
    \item \textit{Limitations of Physical Inspection Techniques:} Conventional inspection techniques, including X-ray imaging, Computed Tomography (CT) scanning, and Explosives Trace Detection (ETD), have inherent limitations in detecting sophisticated threats. These limitations highlight the urgent need for advanced methodologies capable of overcoming these deficiencies.
    \item \textit{Counterfeit and Trojan Insertion:} Counterfeit components and hardware Trojans pose additional challenges, as these can introduce backdoors, degrade performance, or sabotage functionality. Reverse engineering by adversaries facilitates such malicious modifications, while inadequate supply chain verification allows these components to penetrate critical systems undetected. Combatting this requires more rigorous testing and secure design principles.
    \item \textit{Integration of Cyber-Physical Threats:} The growing reliance on embedded systems and battery-powered devices introduces new attack surfaces at the cyber-physical interface. For example, attackers can exploit vulnerabilities in Battery Management Systems (BMS) to manipulate data, induce thermal runaway, or degrade performance. Securing these interfaces requires a holistic approach addressing both physical and cyber vulnerabilities.
\end{enumerate}

\subsection{Emerging Approaches for Detecting Tampered Devices}

\begin{enumerate}
    \item \textit{Advanced Imaging Technologies:} Emerging imaging methods discussed in the previous sections offer promising solutions to overcome the limitations of traditional techniques.

    \item \textit{Blockchain for Supply Chain Traceability:}Blockchain technology offers a decentralized and immutable ledger for tracking components throughout the supply chain. By recording every transaction and movement, blockchain can ensure traceability, authenticate components, and detect tampering at any stage. Coupled with IoT-based sensors, blockchain can provide real-time monitoring and secure the provenance of critical electronics.

    \item \textit{AI-Driven Anomaly Detection:}Artificial intelligence (AI) and machine learning (ML) can enhance the detection of tampered devices by analyzing patterns in physical inspection data, such as X-ray or THz scans. These systems can identify anomalies that human operators might overlook, significantly improving the accuracy and reliability of inspections.

    \item \textit{Tamper-Resistant Design Principles:}Incorporating tamper-evident and tamper-resistant features during the design phase can mitigate risks. Techniques such as physical unclonable functions (PUFs) and secure hardware enclaves can make unauthorized modifications detectable. Additionally, designing systems with self-destruct mechanisms for sensitive applications can prevent exploitation of compromised devices.

    \item \textit{Cyber-Physical Security Frameworks:} Addressing the convergence of cyber and physical vulnerabilities requires integrated security frameworks. These frameworks should combine intrusion detection systems, secure firmware updates, and redundancy mechanisms to protect embedded systems from both physical and cyber threats. For instance, implementing fail-safe mechanisms in BMS architectures can prevent catastrophic outcomes in tampered devices.
\end{enumerate}

\section{Conclusion} \label{sec:Conclusions}
The security concerns surrounding the supply chain of batteries have become increasingly critical, necessitating thorough examination and proactive measures. This paper challenges the traditional belief that devices such as pagers and walkie-talkies are inherently safe from cyberattacks. By analyzing the state-of-the-art attack vectors observed in the Lebanon-based incidents, we have demonstrated the vulnerabilities of these devices and extended the discussion to other battery-powered systems. In addition to identifying the expanding attack surfaces, we explored feasible solutions to address supply chain assurance challenges. Emphasis was placed on advanced physical inspection techniques, including quantum imaging and neutron imaging, which hold promise for enhancing battery security and integrity. With this survey, we aim to draw the attention of both industry and academia to the urgent need for innovative approaches to securing battery technologies. By highlighting vulnerabilities and presenting forward-looking methodologies, we hope to inspire new research directions and foster the development of robust frameworks that ensure the security and resilience of battery-powered systems in the evolving technological landscape.

\bibliographystyle{IEEEtran}
\bibliography{mybibliography}

\end{document}